\documentclass[twoside,onecolumn]{article}

\usepackage{blindtext} 

\usepackage[sc]{mathpazo} 
\usepackage[T1]{fontenc} 
\usepackage{microtype} 

\usepackage[english]{babel} 

\usepackage[hmarginratio=1:1,top=32mm,columnsep=20pt]{geometry} 
\usepackage[hang, small,labelfont=bf,up,textfont=it,up]{caption} 
\usepackage{booktabs} 

\usepackage{lettrine} 

\usepackage{enumitem} 
\setlist[itemize]{noitemsep} 

\usepackage{abstract} 

\usepackage{titlesec} 
\renewcommand\thesection{\Roman{section}} 
\renewcommand\thesubsection{\roman{subsection}} 
\titleformat{\section}[block]{\large\scshape\centering}{\thesection.}{1em}{} 
\titleformat{\subsection}[block]{\large}{\thesubsection.}{1em}{} 

\usepackage{fancyhdr} 
\usepackage{graphics}
\usepackage{graphicx}
\usepackage{epsfig}
\usepackage{amssymb}
\usepackage{cite}
\usepackage{titling} 

\usepackage{hyperref} 

\usepackage{ragged2e}
\usepackage{amsmath}

\pretitle{\begin{center}\Huge\bfseries} 
\posttitle{\end{center}} 
\title{On the collective behavior of interacting polydisperse microbubbles} 
\author{%
\textsc{H. Haghi, AJ. Sojahrood and Michael C. Kolios} \\[1ex]
\normalsize Department of Physics, Ryerson University, Toronto, Ontario, Canada. \\
\normalsize Institute for Biomedical Engineering and Science Technology, \\
\normalsize A Partnership Between Ryerson University and St. Michael’s Hospital, Toronto, Canada. \\
\normalsize Keenan Research Center for Biomedical Science, \\
\normalsize Li Ka Shing Knowledge Institute, St Michael’s Hospital, Toronto, Canada.\\
\normalsize \href{mailto:hossein.haghi@ryerson.ca}{hossein.haghi@ryerson.ca} 
}
\date{\today} 
\renewcommand{\maketitlehookd}{%
\begin{abstract}
 Exploiting the full potential of MBs for applications requires a good understanding of their complex dynamics. Improved understanding of MB oscillations can lead to further enhancement in optimizing their efficacy in many applications and also invent new ones. Previous studies have predominantly addressed the behavior of a single isolated MB in an infinite liquid domain, whereas most applications employ MBs in clusters. Oscillating MBs have been shown to generate secondary pressure waves that modify the dynamics of the MBs in their proximity. A modified Keller-Miksis equation is used to account for inter-bubble interactions. The oscillatory dynamics of each MB within clusters was computed by numerically solving the resulting system of coupled nonlinear second order differential equations. Frequency response analysis and bifurcation diagrams were employed to track the dynamics of interacting MBs. Here we investigate the dynamics of polydisperse MB clusters over a wide range of acoustic and geometric parameters. An emergent collective behavior within bubble clusters was observed whereby individual dynamics of smaller bubbles were suppressed resulting in a collective behavior dominated by the dynamics of the largest MB within the cluster. The emergent dynamics of smaller MBs within bubble clusters can be characterized by constructive and destructive inter-bubble interactions. In constructive interactions, the radial oscillations of smaller bubbles matched those of the largest MB and their oscillations are amplified. In destructive interactions, the oscillations of smaller bubbles are suppressed so that their oscillations match those of the largest MB. Furthermore, a special case of constructive interactions is presented where dominant MB (largest) can force smaller MBs into period doubling and subharmonic oscillations.
\end{abstract}
}

\begin{document}

\maketitle


\section{Introduction}
\justifying
Ultrasonically excited MBs have shown to exhibit highly nonlinear and complex dynamics \cite{1,2,3,4,5,6,7,8,9,10,11,12,13,14,15,16,17,18,19,20}. Due to their high echogenicity, MBs are employed in ultrasound imaging as contrast agents \cite{21,22}. In context of ultrasound imaging, nonlinear subharmonic (SH) and ultraharmonic (UH) response of MBs is exploited in diagnostic ultrasound to enhance detection of vascular tissues \cite{23,24,25} since tissue generates negligible amounts of subharmonics and ultraharmonics in response to ultrasonic waves \cite{24,26}. SH oscillations of the microbubbles may also be employed for non-invasive pressure estimation in medical ultrasound \cite{27,28}. This can generate significant suppression of tissue signal that results in high sensitivity contrast enhanced imaging. Using certain ultrasound exposure parameters \cite{29}, MBs can undergo inertial collapse generating powerful shockwaves and high velocity jets which are exploited in shockwave lithotripsy \cite{30} and histotripsy \cite{31}.Oscillations of MBs can also generate microstreaming in the surrounding liquid \cite{32}, exerting shear stress on the nearby objects \cite{33} allowing for therapeutic approaches such as site specific drug and gene delivery \cite{34} through the creation of pores on the cell membrane \cite{35}. Similar mechanisms are implicated in the reversible opening of the blood-brain-barrier (BBB) to deliver macromolecules to better treat central nervous system disorders and brain cancers \cite{37,38}; SH response of the bubbles have been proposed as a marker for monitoring the treatment \cite{39} BBB opening \cite{40,41} and enhancement of thrombolysis \cite{42}. MBs are involved in sonoluminescence, further enhancing the efficacy of chemical reactions in sonochemical reactors \cite{43,44}. In microfluidic applications, the response of MBs to an ultrasonic wave has been utilized to stir or pump liquids on miniature scales \cite{45}. MBs are also of interest in underwater acoustics \cite{46} and oceanography \cite{48}.

Understanding the complex dynamic of ultrasonically driven MBs is the key to optimize their performance in applications ranging from underwater acoustics to medicine.

There have been decades of theoretical and experimental investigation on the dynamics of MBs under acoustic excitation. Most of these studies are based on the assumption that MBs within a cluster are sufficiently apart from each other so that their oscillations are independent. These investigations have provided valuable insight on the dynamic behavior of acoustically excited isolated individual MBs in an infinite liquid domain. Nevertheless, the majority of MB applications involve dense clusters of polydisperse MBs \cite{48,49,50}.
Oscillations of each MB within a bubble cluster generates secondary pressure waves \cite{50}, exerting pressure on the neighboring MBs. Therefore, each bubble oscillates in response to two main pressure fields, the primary acoustic ultrasound pressure wave and secondary pressure waves scattered from all the bubbles in proximity. Hence, a MB cluster can be considered as a system of coupled oscillators where the dynamics of each oscillator depends on the rest \cite{51}.

 It has been shown that when MBs are close to each other the effects of coupling can be significant. For instance, previous investigations have concluded that independent oscillations of MBs are suppressed due to inter-bubble interactions and therefore a collective behavior within a bubble cluster emerges \cite{51}. Moreover, it has been shown that inter-bubble interactions can increase the maximum radius of MB oscillation \cite{52}. Another study has concluded that route to chaos could be altered by inter-bubble interactions where suppression of chaotic oscillations is possible by changing cluster size \cite{53}. It was shown in the same study that larger MBs have a significant impact on the oscillations of smaller MBs. Pioneering work of Allen et al. \cite{54} on the interaction of two ultrasound contrast agents showed that the total radiated pressure from the bubbles depends on the extent of coupling, which is a function of the separation distance of the two agents and during in vivo experimental and clinical conditions, coupling becomes an important issue.

 Although these studies have highlighted the importance of inter-bubble interactions, they lack a general framework by which one can predict the emergent collective behavior within MB clusters.
This work aims to address this by studying the effects of MB interactions in polydisperse MB clusters.  The aim of this work is to provide a framework where one can predict the emergent collective dynamics of interacting MBs within a cluster. Inter-bubble interactions are classified into two general categories characterized by destructive and constructive interactions. Using the proposed classification of inter-bubble interactions, one can predict the emergent collective behavior of MB clusters.

In this work, a set of detailed numerical simulations for a wide range of geometric and acoustic parameters are performed. The resulting time dependent radius of oscillators were studied through resonance and bifurcation analysis techniques. The emergent behavior of isolated MBs when they interact within a cluster can be predicted using their frequency response diagrams. Two general trends from the study of the frequency response diagrams of isolated and interacting MBs are observed. In the first trend, the resonance modes of smaller MBs are enhanced (or new resonances are formed) upon interaction with other MBs (constructive interaction). In the second trend, a weakening (or disappearance) of resonance modes are observed upon interaction with other MBs (destructive interaction). Furthermore, the contribution of smaller and larger MBs to the emergent collective behavior of polydisperse MB clusters is investigated. A special case of constructive interaction where the largest MB within the cluster can force smaller MBs into period two oscillations is studied in detail with the aid of frequency dependent bifurcation diagrams.

\section{Methods}
\subsection{The Bubble Model}
\justifying
Dynamics of a single isolated bubble in an infinite domain of liquid is governed by the well-known Keller-Miksis equation (Eq.1) which assumes small Mach numbers for the oscillating bubble wall \cite{55}. The Keller-Miksis equation is a highly nonlinear second order ordinary differential equation given by

	\begin{equation}
	\rho{}\left(1-\frac{\dot{R}}{c}\right)R\ddot{R}+\frac{3}{2}\rho{}{\dot{R}}^2\left(1-\frac{\dot{R}}{3c}\right)=\frac{P_B(R,\dot{R},t)}{\rho_l}
	\end{equation}
	
Where the nonlinear term $P_B$ can be written as:
	\begin{equation}
	 P_B=[1+\frac{\dot{R}}{c}+\frac{R}{c}\frac{d}{dt}][(P_\infty+\frac{2\sigma{}}{R_0}){(\frac{R_0}{R})}^{3k}-\frac{2\sigma{}}{R}-\frac{4\mu_L{}\dot{R}}{R}-P_\infty+P_asin\left(2\pi{}ft\right)]
	\end{equation}

In Eq.1 and Eq.2 $R$, $\dot{R}$, $\ddot{R}$ respectively are instantaneous radius, wall velocity and wall acceleration of the bubble. $R_0$ is the initial radius of the bubble and $\sigma$, $\mu_L$ and $c$ respectively are the surface tension, viscosity and speed of sound in the liquid. $\gamma$ is the ratio of specific heats of the gas within the bubble, $P_\infty$ is the ambient pressure in the surrounding liquid, $P_a$ and $f$ respectively are the pressure amplitude and frequency of the incoming acoustic field.
	
Successive expansions and contractions of an oscillating bubble in an incompressible fluid generates pressures waves whose amplitude can be calculated using Euler's equation of fluid flow as:
	\begin{equation}
	P_{sc}=\frac{\rho_L}{r}\frac{d(R^2 \dot{R})}{dt}+o(\frac{1}{r^4})
	\end{equation}
	Neglecting higher orders terms in Eq.3 leads to Eq.4 which was suggested in \cite{56}
	\begin{equation}
	P_{sc}=\rho_L\frac{2R\dot{R}^2+\ddot{R}R^2}{r}
	\end{equation}
where $r$ denotes the distance of the point of interest from center of the bubble. In Eq.1, the term $P_B$ refers to the pressure on the bubble wall. In a cluster of multiple MBs, the pressure at the wall of each bubble is equal to $P_B$ and the summation of all scattered pressure waves from other bubbles in the cluster. Therefore, coupling between MBs is done through inclusion of backscattered pressure fields from all the MBs in the cluster at the location of each MB. Thus, the Keller-Miksis equation for a cluster of interacting bubbles can be written as
	
	\begin{equation}
	\left[1-\frac{\dot{R_i}}{c}\right]R_i\ddot{R_i}+\frac{3}{2}\left[1-\frac{\dot{R_i}}{3c}\right]\dot{R_i^2}=\frac{P_B(R_i,\dot{R_i},t)}{\rho_L}-\sum_{j=1,j\neq i}^{N} \frac{2R_j\dot{R_j}^2+\ddot{R_j}R_j^2}{d_{ij}}
	\end{equation}
where $R_i$, $\dot{R_i}$ and $\ddot{R_i}$ respectively represent instantaneous radius, wall velocity and wall acceleration of each bubble. $d_{ij}$ represents the distance between centers of $i$th and $j$th bubble. $R_{i0}$ is the initial radius of the $i$th bubble and $N$ is the number of bubbles in the cluster.
	
Eq.5 is a system of $N$ linearly coupled, second order differential equations; a solution requires $2N$ initial conditions. Initial conditions can be specified by assuming all bubbles are initially at rest (no radial velocity) and that their initial radii are known. The initial conditions can be written as:
	\begin{equation}
	\begin{cases}
	R_i(t=0)=R_{i0}\\
	\dot{R_i}(t=0)=0
	\end{cases}   i=1,2,...,N
	\end{equation}
\justifying	
Therefore, Eq.5 can be treated as an initial value problem for which the Runge-Kutta numerical algorithms can be employed to obtain a solution. Furthermore, the geometry of the problem needs be specified to determine the MB distances ($d_{ij}$).
	
The aim of this study is to investigate nonlinear effects of inter-bubble interactions, therefore, additional nonlinearities such as the effects of the encapsulating shell \cite{57,58} are not introduced in this study. To further reduce the ambiguity in the source of the observed effects, spatial formations were chosen in which the inter-bubble distances between all MBs are equal ($d_{ij}=d$). Therefore, equilateral tetrahedrons (4 MBs) and equilateral triangles (3 MBs) were chosen to specify the geometry of the problem (Fig.1).
	
	\begin{figure*}
		\begin{center}
			\scalebox{0.2}{\includegraphics{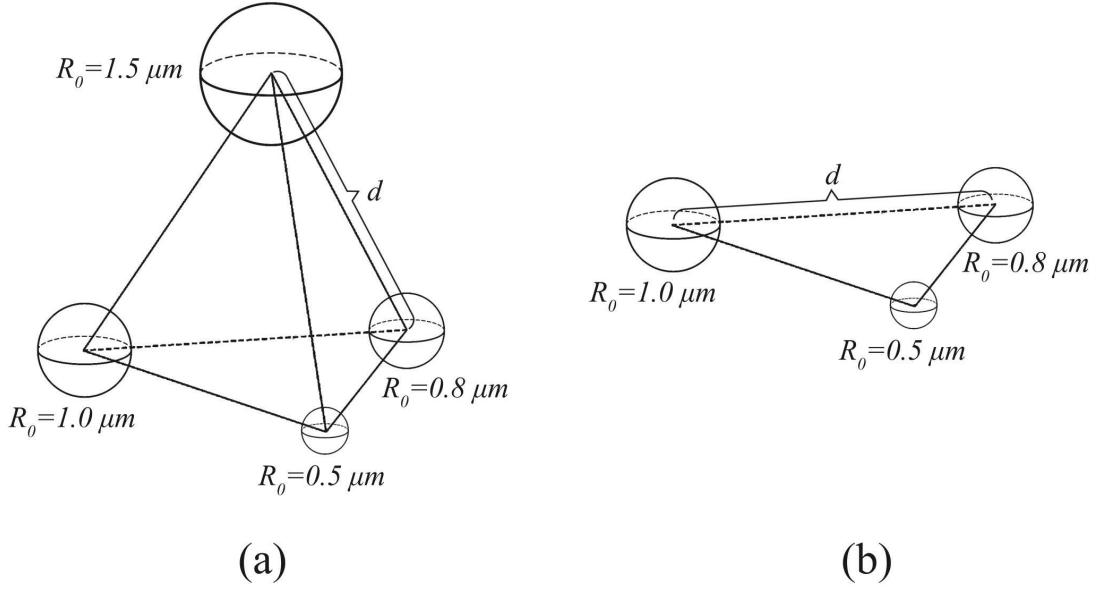}}
			\caption{Schematic diagrams of the (a) equilateral tetrahedron formation for the 4-MB cases, (b) equilateral triangle formation for the 3-MB case}
		\end{center}
	\end{figure*}
	
	\subsection{Numerical analysis methods}
	\justifying
	Since analytical solutions are not available for Eq.5, numerical methods are employed to solve the equations. Eq.5 should be solved for a wide range of the parameters of interest to evaluate the impact of these parameters on the dynamics of the system. Bifurcation and frequency response analysis were used to investigate the behavior of this complex system for a wide range of control parameters. These methods have been pioneered by Lauterborn and Parlitz \cite{38,39}. Here, a brief description of the bifurcation and frequency response analysis is presented.
	
	\subsubsection{Bifurcation analysis}
	\justifying
	Stroboscopic maps based on mapping of radius of MBs after each forcing period have been used to evaluate the dynamics of MBs for a wide range of control parameters \cite{1,4,8,61,62}. After the system reaches its steady state, the MB radii are sampled and plotted as a function of the controlling parameter in a Poincare plot. The procedure continues by modifying the control parameter and obtaining a new set of points to be plotted against the control parameter. For instance, here we consider a MB with an initial radius of 2$\mu$m sonicated with ultrasound waves of with a frequency of 2MHz and pressures ranging from 1kPa up to 400kPa. A solution to the time dependent radial oscillations of the MB at lower pressures is show in Fig.2a. Oscillations are repeated after each forcing period resulting in the same $R/R_0$ values for the sampled radius after each forcing period (Period-1 oscillations). The sampled points (after the transient phase) from Fig.2a results in a single point in Fig.2d for a particular set of exposure parameters. Increasing the incident pressure increases $R/R_0$, and above a certain pressure threshold the time-radius curves change shape (Fig.2b). Periodic sampling of the MB radius results in two values (Period-2 oscillations). Plotting the sampled points on a Poincare diagram as a function of pressure amplitude results in two points as shown in Fig.2d. Further increasing of pressure above a specific threshold results in chaotic oscillations depicted in Fig.2c where sampling of the points results in many different values (Chaotic oscillations). The chaotic oscillations generate multiple points on the bifurcation diagrams presented in Fig.2d. Similar analysis can be performed to investigate the influence of any control parameter.

	\begin{figure*}
		\begin{center}
			\scalebox{0.15}{\includegraphics{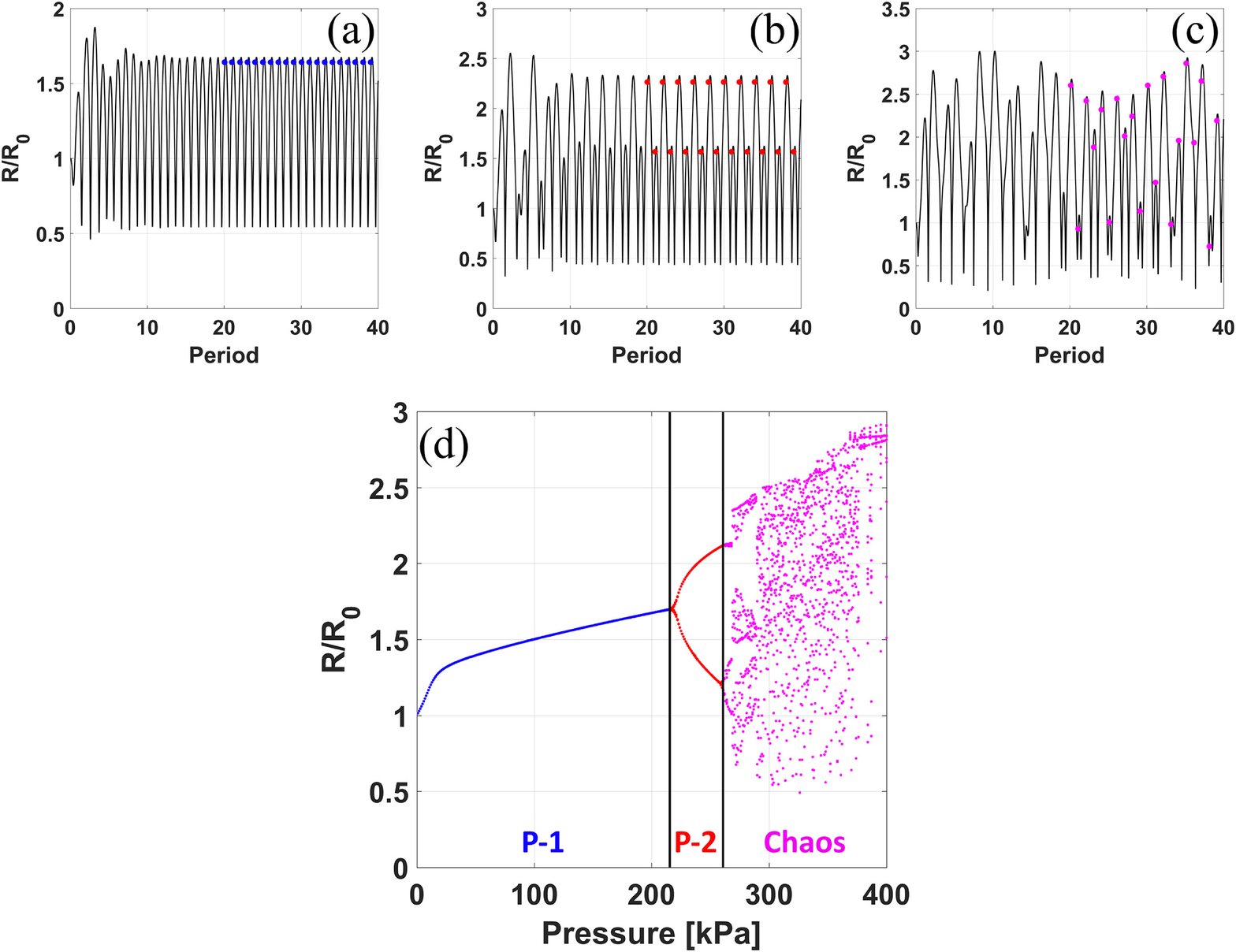}}
			\caption{Radial Oscillations of a MB with initial radius of 2$\mu$m exposed to an ultrasound wave with a frequency of 2MHz, (a) Peroid-1 oscillations ($P_a$=100 kPa), (b) Period-2 oscillations ($P_a$=250 kPa), (c) Chaotic oscillations ($P_a$=350 kPa), (d) Pressure dependent bifurcation diagram of a 2$\mu$m MB excited with an ultrasound wave with a frequency of 2MHz.}
		\end{center}
	\end{figure*}
	
	\subsubsection{Frequency response analysis}
	\justifying
	Frequency response graphs characterize the MB radial oscillation amplitude at a fixed pressure as a function ultrasound of frequency. The graphs can be used to determine the MB resonance frequencies and their dependence on various parameters \cite{55,59} . They are generated through the following steps:
	\begin{enumerate}
		\item Radial MB oscillations are generated using an ultrasound wave with a fixed pressure amplitude using multiple frequencies ranging from $f_{min}$ to $f_{max}$ with steps of $f_{step}$.
		\item The steady state portion of the oscillations is examined for a range of ultrasound frequencies.
		\item The maximum $R/R_0$ value within each selected window is plotted as a function of frequency in a separate graph
	\end{enumerate}
	
	Fig.3 illustrates this process for a MB of 2$\mu$m initial radius excited at a pressure amplitude of 1kPa using frequencies ranging from 0.5MHz up to 5MHz with steps of 0.01MHz. Oscillations at these low pressures are termed as linear \cite{63} where $\mid R/R_0\mid<<1 $. In linear oscillation regimes, the frequency response graph has only one peak that corresponds to linear resonance frequency of the MB (Fig.3d).
	
	At higher pressure amplitudes, secondary peaks (subharmonics, higher harmonics, etc.) will appear on the frequency response graph which was shown and named previously \cite{59}.
	
	\begin{figure*}
		\begin{center}
			\scalebox{0.21}{\includegraphics{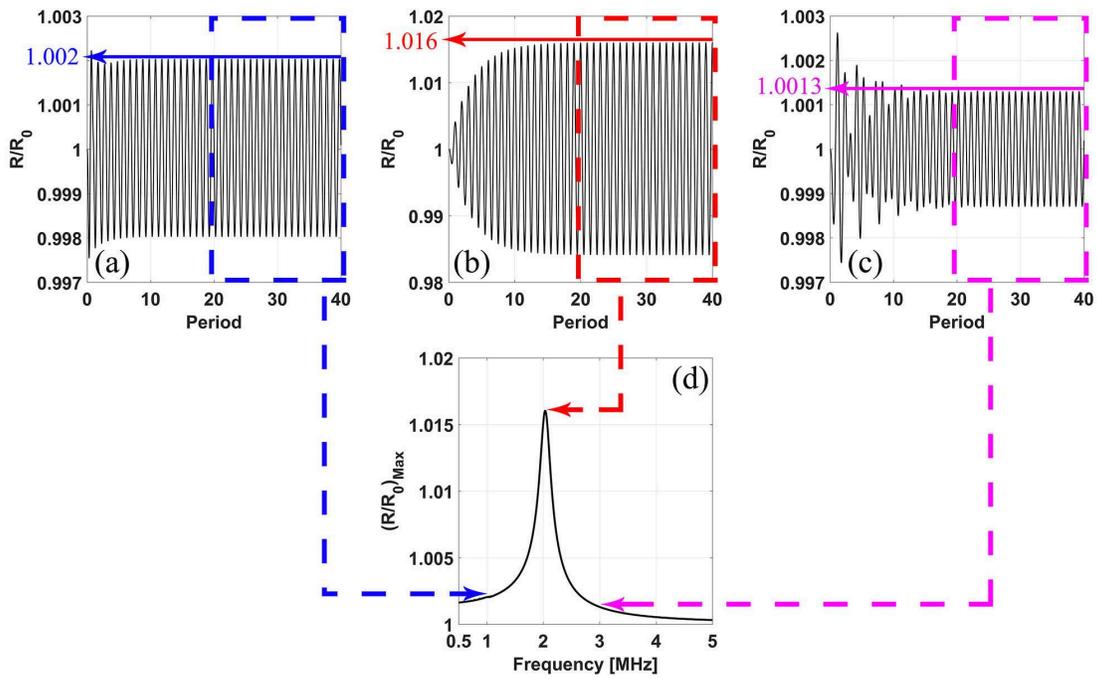}}
			\caption{Radial oscillations of a MB with initial radius of 2$\mu$m excited with ultrasound wave of 1kPa pressure amplitude (a) below resonance oscillations (f=1MHz), (b) Resonant oscillations (f=2.04 MHz), (c) Above resonance oscillations (f=3MHz). (d) Frequency response graph of a MB with initial radius of 2$\mu$m excited with ultrasound wave of 1kPa pressure amplitude}
		\end{center}
	\end{figure*}
	
	\subsection{Parameters}
	\justifying
	Two main classes of 3 microbubbles (3-MB) and 4 microbubbles (4-MB) clusters are simulated under a wide range of ultrasonic and geometric scenarios. Initial radii of MBs are chosen as 0.5$\mu$m, 0.8$\mu$m, 1.0$\mu$m and 1.5$\mu$m to represent a size range of interest in medical applications \cite{64,65}. 4-MB clusters are formed by placing centers of the MBs on the vertices of an equilateral tetrahedron of side $d$ illustrated in Fig.1a. 3-MB clusters are then assembled with removal of the largest MB from the 4-MB cluster, resulting in a equilateral triangle spatial formation. This allows the examination of the influence of the largest MB within the cluster on the dynamics of smaller MBs and consequently the collective behaviour of the MB cluster. In the 4-MB cluster the largest MB has a 1.5$\mu$m initial radius and in the 3-MB cluster the largest MB has a 1.0$\mu$m initial radius. $d$s represent the equal distance between MBs, ranging from 5$\mu$m to 300$\mu$m for the largest separation distance. Simulation results were generated as a function of distance between the MBs in steps of 0.1$\mu$m from 5$\mu$m up to 20$\mu$m and in steps of 0.5$\mu$m for larger distances. 60 ultrasound pulses with pressure amplitudes of 120kPa and frequencies ranging from 0.5MHz up to 15MHz with 3kHz increments were utilized to excite the MBs. The pressure amplitude is chosen such that only the largest chosen MB ($R_0=1.5\mu m$) can exhibit subharmonic resonance in isolation in the chosen frequency range.
	
	\section{Results}
	\justifying
	
	The frequency response and bifurcation diagrams of isolated and interacting MBs were compared.  Fig.4a illustrates the frequency response for each of the MBs in the 4-MB cluster without interaction ($d\to\infty$). Fig.4b depicts the frequency response graph of interacting MBs at a separation distance of 5$\mu$m. Removal of the largest MB in the 4-MB cluster results in the frequency response graph shown in Fig.4c. The interaction of MBs significantly change the MB oscillation dynamics.In the following section details of Fig.4 are discussed.
	
	\begin{figure*}
		\begin{center}
			\scalebox{0.45}{\includegraphics{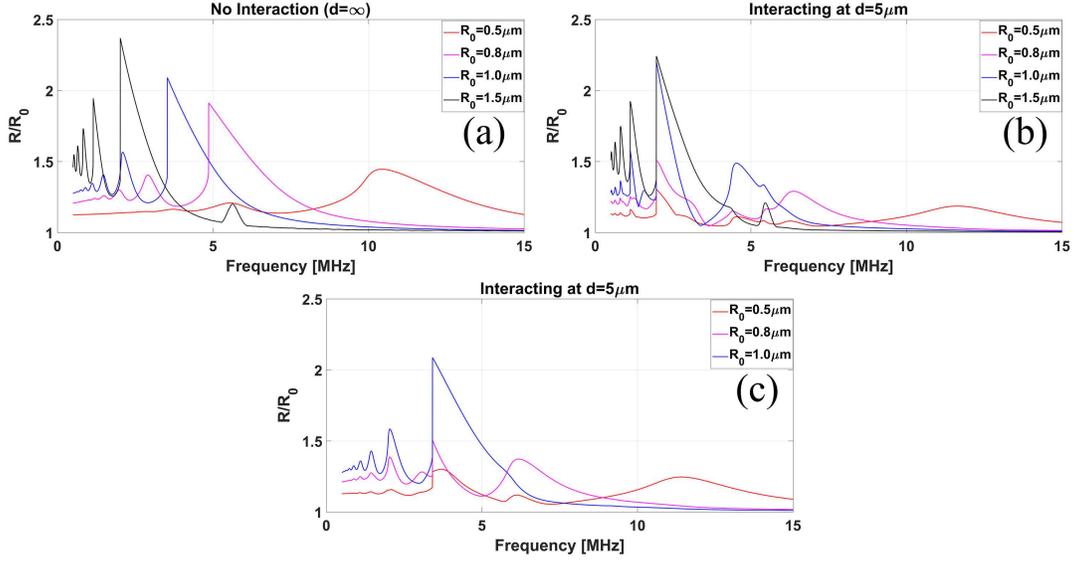}}
			\caption{Frequency response graph of non-interacting 4-MB (a), Interacting 4-MB at $d=$5$\mu$m (b), interacting 3-MB at d=5$\mu$m using an ultrasound wave of $P_a$= 120kPa with frequencies ranging from 0.5MHz up to 15MHz}
		\end{center}
	\end{figure*}

	\subsection{Resonance formation, enhancement and suppression}
	\justifying
	Fig.5a and Fig.5b respectively present the frequency response diagrams of isolated non-interacting and a cluster of interacting MBs at a separation distance of 5$\mu$m presented in Fig. 4, focused on the range of frequencies between 0.5-3MHz. In Fig.5a, the 1/1 harmonic resonance peak of the 1.5$\mu$m MB (black arrow) is aligned (similar frequencies) with the 2/1 harmonic resonance peak of the 1.0$\mu$m MB (blue arrow) as well as the 3/1 harmonic resonance peak of the 0.8$\mu$m MB (magenta arrow). Moreover, Fig.5a shows that the 2/1 harmonic resonance mode of the largest MB (orange arrow) is aligned with 4/1 harmonic mode of the 1.0$\mu$m MB (green arrow). Furthermore, Fig.5a shows that the smallest isolated MB does not have any resonance modes in frequencies lower than 3.5MHz and its 1/1 harmonic resonance mode occurs at approximately 11 MHz (Fig.4a).

	\begin{figure*}
		\begin{center}
			\scalebox{0.85}{\includegraphics{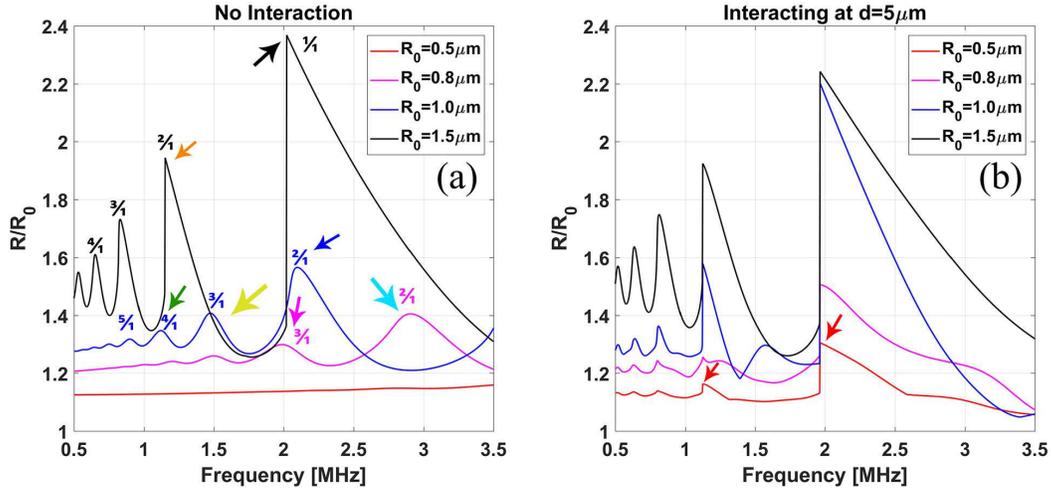}}
			\caption{Frequency response graph of (a) non-interacting 4-MB cluster (b) Interacting 4-MB cluster at $d$=5$\mu$m using an ultrasound wave of $P_a$=120kPa with frequencies ranging from 0.5MHz up to 3.5MHz}
		\end{center}
	\end{figure*}

	Fig.5b shows how bubble interactions change the oscillation dynamics. Fig.5b demonstrates a significant enhancement in the aforementioned aligned peaks in Fig.5a and also the formation of new resonance modes for the 0.5$\mu$m MB (red arrows). There is a negligible weakening in the resonance amplitudes of the largest MB. Moreover, Fig.5a shows that all of the resonance modes of the smaller MBs that are not aligned with any of the resonance modes of the largest MB are suppressed. This is shown in Fig.5b for which MBs are interacting. For instance, the 3/1 harmonic resonance mode of the 1$\mu$m MB (yellow arrow) and 2/1 harmonic resonance mode of the 0.8$\mu$m MB (cyan arrow) in Fig.5a, as well as the 1/1 harmonic resonance mode of the 0.5$\mu$m MB in Fig.4a, are suppressed significantly when MBs are interacting.
	
	Next, the largest MB ($R_0=1.5\mu m$) from 4-MB clusters is removed to form a 3-MB clusters. The resonance analysis with control parameters equal to the ones used in 4-MB cluster case were performed. The resulting frequency response diagrams for the non-interacting 3-MB cluster and interacting 3-MB cluster are illustrated respectively in Fig.6a and Fig.6b.
	
	\begin{figure*}
		\begin{center}
			\scalebox{0.85}{\includegraphics{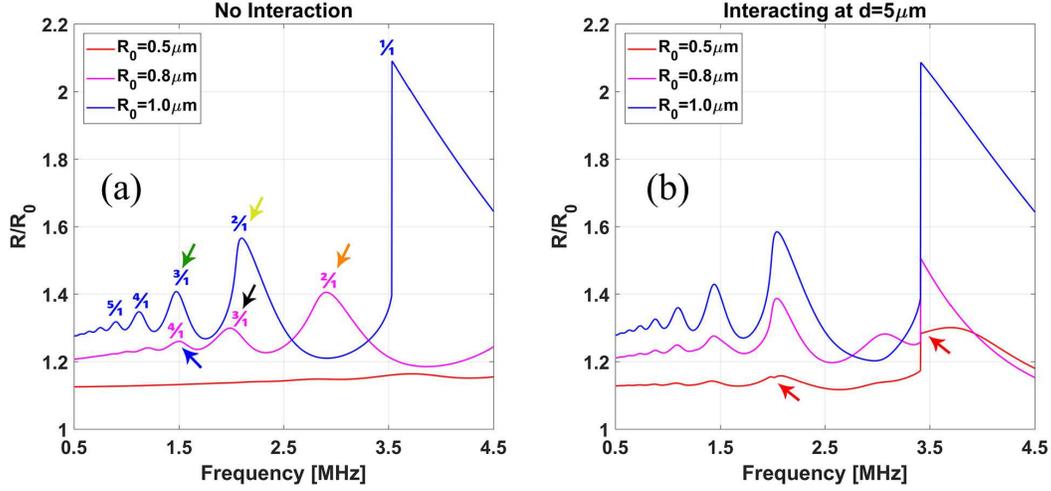}}
			\caption{Frequency response graph of (a) non-interacting 3-MB cluster (b) Interacting 3-MB cluster at $d$=5$\mu$m using an ultrasound wave of $P_a$=120kPa with frequencies ranging from 0.5MHz up to 4.5MHz}
		\end{center}
	\end{figure*}	
	
	A trend similar to the one observed in the 4-MB case is also present in the 3-MB case. In the 3-MB configuration, the largest MB in the cluster has an initial radius of 1$\mu$m. Fig.6a shows that the 3/1 (black arrow) and 4/1 (yellow arrow) harmonic mode of the 0.8$\mu$m MB (which are aligned respectively with 2/1 (red arrow) and 3/1 (green arrow) harmonic modes of the 1.0$\mu$m MB) are enhanced in Fig.6b in which the MBs are interacting. Moreover, similar to the 4-MB case, Fig.6a shows that 2/1 harmonic (orange arrow) resonance peak of the 0.8$\mu$m MB (not aligned with any of the peaks of the largest MB) is weakened in Fig.6b when the MBs are interacting. Formation of secondary new peaks due to the MB interaction is also evident in Fig.6b, similar to the ones formed under the 1/1 harmonic resonance mode of the largest MB (red arrows).
	
	\subsection{Forced subharmonic resonance and period doubling}
	Fig.7 illustrates the frequency response curve for all three configurations when $P_a=120kPa$. Fig.7a shows that in the absence of interaction, only the largest MB ($R_0=1.5\mu m$) exhibits a 1/2 subharmonic resonance peak (black arrow). Fig.7b illustrates that at an inter-bubble distance of 5$\mu m$, new resonance peaks in the frequency response of the smaller MBs (red, magenta and blue arrows) are formed. Removal of the largest MB from the 4-MB cluster (Fig. 7c) results in elimination of the new resonance modes of the smaller MBs in the examined frequency range. This suggests that the largest MB in the cluster is responsible for the formation of new resonance peaks in the frequency response curve of the smaller MBs. The new resonance peak results in period-2 (SH) oscillations in all the bubbles within the examined cluster.
	
	\begin{figure*}
		\begin{center}
			\scalebox{1.2}{\includegraphics{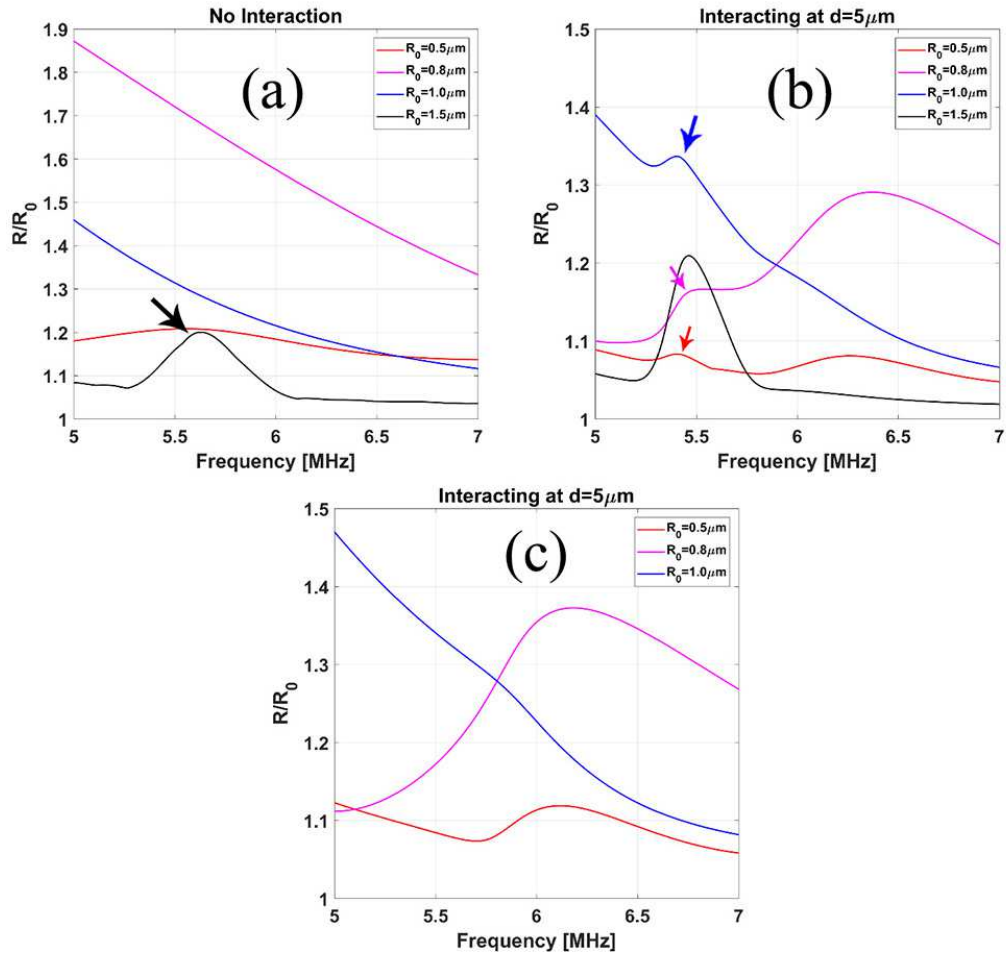}}
			\caption{Frequency response graph of non-interacting 4-MBs (a), Interacting 4-MB at $d$=5$\mu$m (b), interacting 3-MB at $d$=5$\mu$m using an ultrasound wave of $P_a$=120kPa with frequencies ranging from 5MHz up to 7MHz}
		\end{center}
	\end{figure*}	
	
	To better examine the nature of the resonance behaviors in Fig.7, the frequency dependent bifurcation structure of the MBs (Fig.8) in all three configurations (5-6.2 MHz) were generated. Fig.8a shows that in the absence of interaction, only the largest MB with an initial radius of $1.5\mu m$ exhibits P-2 oscillations. However, due to interaction all MBs in the 4-MB cluster perform P-2 oscillations (Fig.8b), in other words the largest MB within the cluster forces the smaller MBs to exhibit P-2 oscillations. Removing the largest MB from the 4-MB cluster reshaped the P-2 oscillations back into P-1 (Fig.8c) indicating the role of the largest MB in forcing smaller MBs to oscillate in P-2 mode.
	
	\begin{figure*}
		\begin{center}
			\scalebox{0.85}{\includegraphics{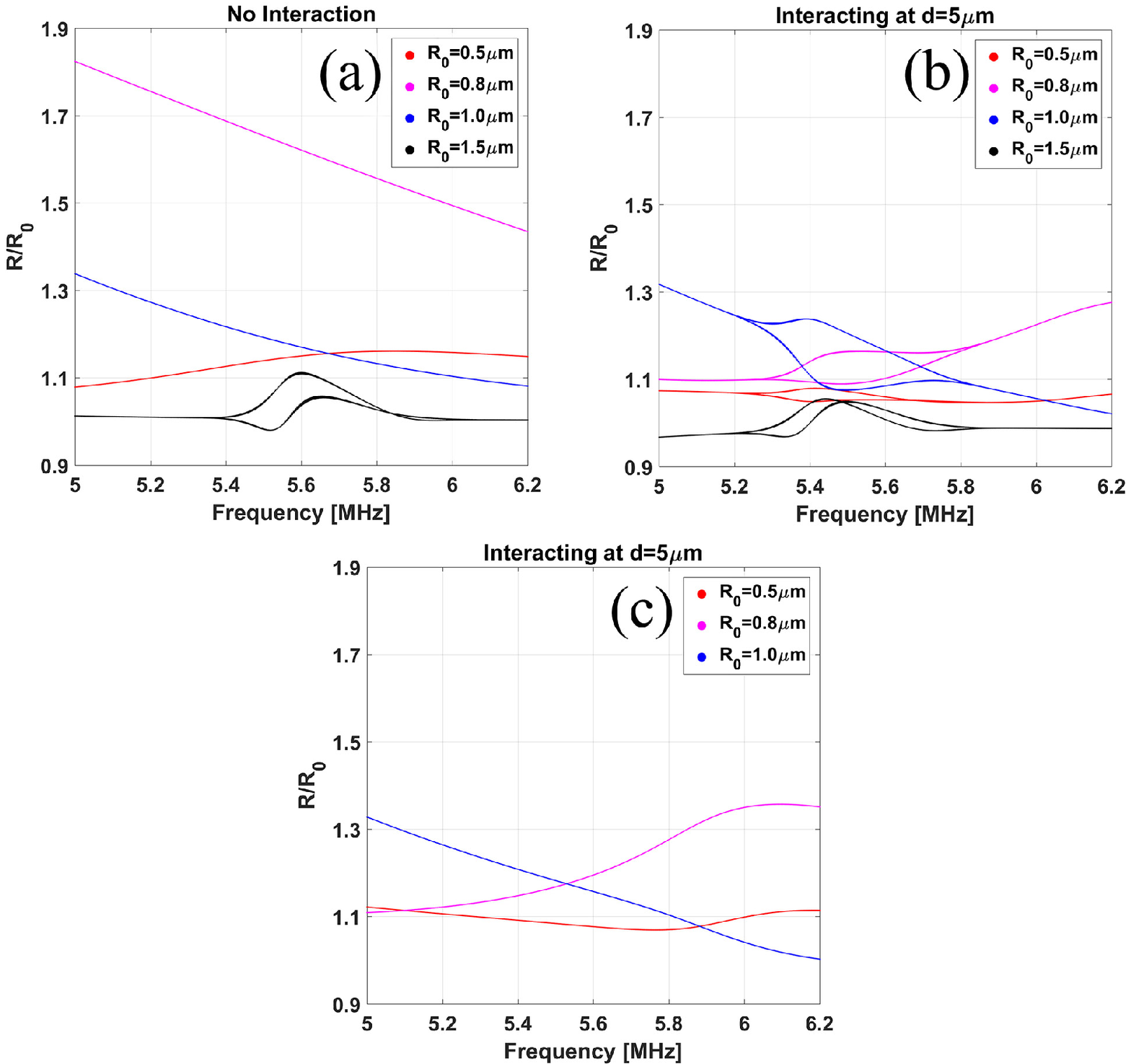}}
			\caption{Frequency dependent bifurcation structure of non-interacting 4-MBs (a), Interacting 4-MB at $d$=5$\mu$m (b), interacting 3-MB (removal of largest MB) at $d$=5$\mu$m using an ultrasound wave of $P_a$=120kPa with frequencies ranging from 5MHz up to 6.2MHz }
		\end{center}
	\end{figure*}	
	
	Further investigation into the forced P-2 oscillations was performed with bifurcation analysis as a function of inter-bubble distance. An ultrasound wave with 120 kPa of pressure amplitude and frequency of 5.4MHz was used. For this exposure parameters, all of the MBs in Fig.8b undergo period doubling when the separation distance is set at $5.0\mu m$.
	
	\begin{figure*}
		\begin{center}
			\scalebox{0.85}{\includegraphics{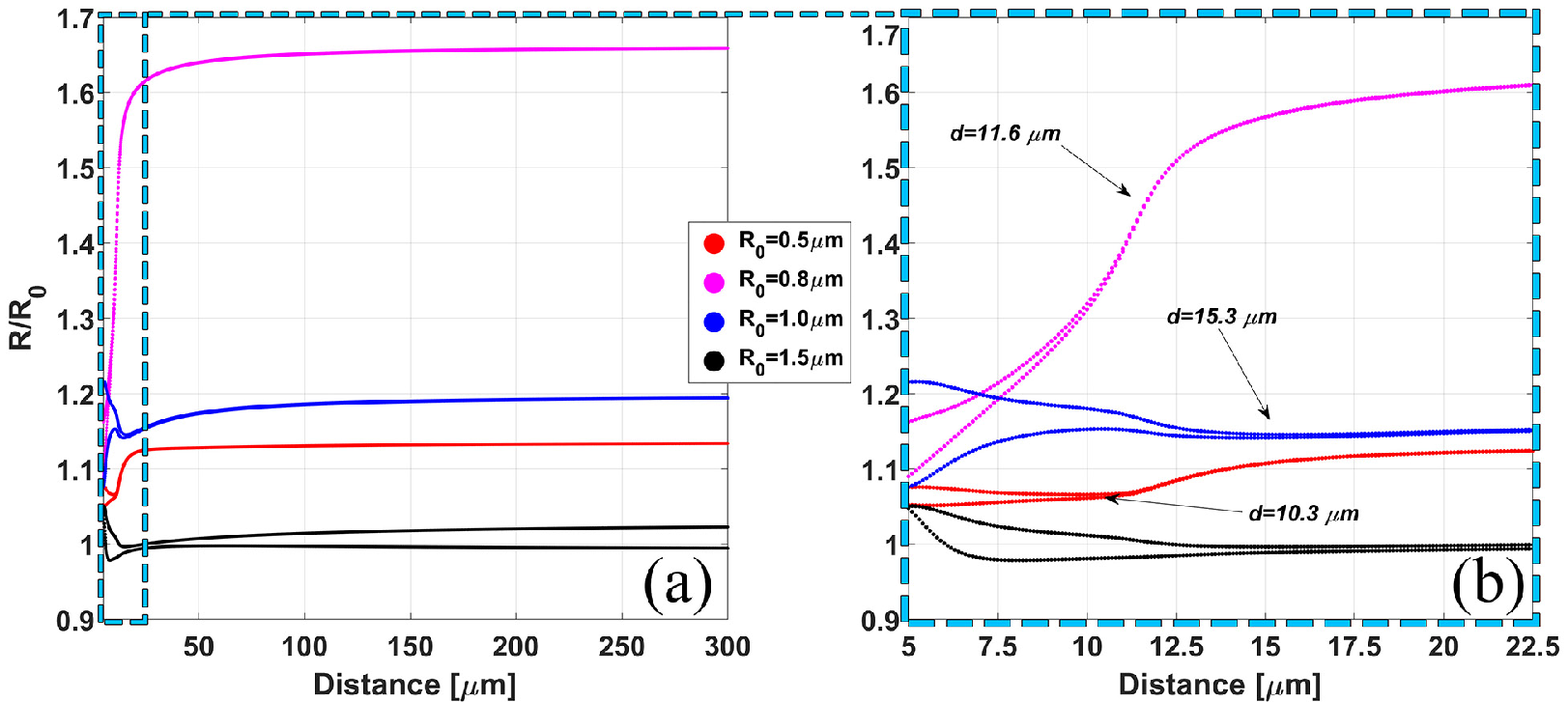}}
			\caption{Inter-bubble distance dependent bifurcation diagram of the 4-MB cluster excited with an ultrasound wave with a frequency of 5.4 MHz and pressure ampltiude of 120 kPa (a) $d$= 5$\mu$m - 300$\mu$m (b) $d$= 5$\mu$m – 22.5$\mu$m}
		\end{center}
	\end{figure*}

	Fig.9 displays the bifurcation structure of the 4 bubbles as a function of the separation distance when $f=$5.4 MHz and $P_a=$120 kPa.  Fig.9a shows that if MBs are sufficiently far apart from each other, none of the smaller MBs undergo period doubling. At closer inter-bubble distances, all of the smaller MBs undergo period doubling. The closer the size of smaller MB to the largest MB, the shorter the distance required for the forced oscillation to occur. The $1.0\mu m$ MB was forced into period doubling at a separation distance of $15.3\mu m$. At this distance, the rest of MB within the cluster are maintaining their P-1 mode. Further reducing the distance between the MBs results in period doubling of the $0.8\mu m$ and $0.5\mu m$ MBs at distances of $11.6\mu m$ and $10.3\mu m$, respectively.
	
	\section{Discussion and summary}
	\justifying
	In this study, we investigated the effects of coupling and inter-bubble interactions on the collective behavior of polydisperse MB clusters. A system of coupled Keller-Miksis equations were used to assess the dynamics of individual interacting MBs within a MB cluster. Coupling was achieved through the inclusion of the backscattered pressure from all the MBs within the cluster at the location of each MB. The resulting system of coupled second order differential equations were solved by defining initial conditions and the spatial geometry of the clusters as equilateral tetrahedron (4 MBs) and equilateral triangles (3 MBs). Numerical solutions of the mathematical model were obtained through a wide range of ultrasonic frequencies and geometric configurations that were solved employing a Runge-Kutta algorithm. The time dependent radial oscillations of each MB were recorded and analyzed with the aid of bifurcation and frequency response analysis techniques.

	Our findings further validate previous observations affirming the significance of the largest MBs and their influence on the smaller MBs within a cluster while remaining mostly unaffected by the oscillations of the smaller MBs. Furthermore, our findings build upon previous observations and suggest a pattern of destructive and constructive interactions within the cluster mainly dictated by the largest MB. We have shown that smaller MBs within the cluster are forced to reshape their oscillations through constructive and destructive interactions to match the largest MB within the cluster. In this regard the important findings can be summarized as follows:

	\begin{enumerate}
		\item Destructive inter-bubble interactions occur if a resonance mode of a smaller isolated MB is not aligned with a resonance mode of the largest isolated MB within the cluster. This results in significant suppression of the non-aligned resonance mode of the smaller MBs. 	
		
		\item Constructive inter-bubble interactions occur if a resonance mode of a smaller isolated MB is aligned with a resonance mode of the largest isolated MB within the cluster. The aligned resonance modes of the smaller MBs are enhanced due to constructive inter-bubble interactions. 	
		
		\item Smaller MBs can be forced to generate subharmonic resonance modes due to constructive inter-bubble interactions in a lower frequency than their natural subharmonic resonance mode ($~ 2 \times $natural resonance frequency \cite{66}). Following the subharmonic resonance generation, the largest MB within the cluster forced smaller MBs to reshape their oscillations from P-1 into P-2 oscillations.
		
		\item The closer the size of the smaller MB to the largest MB within the cluster, the larger the distance for which it can be forced into P-2 oscillations. Smaller MBs require closer proximity to the largest MB to be forced into P-2 oscillations.
	\end{enumerate}
	
	The forced oscillations would amplify or dictate the mode of the oscillatory behavior of the smaller MBs in the MB clusters which thereby controls the behavior of the system. In other words, by controlling the behavior of the largest MBs in the population one may control the behavior of the whole system.  For high concentration MB solutions, for which the distance between the MBs are small enough for strong interaction, the larger MBs could amplify the oscillatory behavior of the smaller MBs, thereby amplifying their effect in imaging or therapy applications. Our numerical calculation showed that the smaller bubbles in the cluster were forced by the bigger bubble to exhibit P-2 oscillations when the separation distance was below $15.3\mu m$ and below the separation distance of $10.3\mu m$ all of the three smaller bubbles were forced by the bigger bubble to exhibit P-2 and SH oscillations. Allen et al.\cite{54} estimated the separation distances between the bubbles in a real clinical application. He considered a typical $3cc$ injection of MBs into $5L$ human blood and estimated an average separation distance of a $14.4\mu m$ and $8.4\mu m$ for monodisperse populations of $1\mu m$ radius bubbles representative of the concentrations that is used in clinical applications. Thus, our findings show that interaction will become significant in clinical applications of MBs and the behavior of the cluster needs to be optimized by taking into account the interaction.
	
	Findings of this study may provide basic fundamental framework to optimize the behavior of a polydisperse population of MBs. As an example, the concentration and the size distribution of the population maybe engineered to enhance SH oscillations. This effect can be particularly  important in the case of nanobubbles, for which higher concentrations are typically used \cite{67,68,69}. Another example is to choose exposure parameters (frequency and pressure) by which the destructive interference is avoided; this will result in a signal with higher strength and may enhance the signal to noise ratio in imaging applications or enhancing the streaming velocities and the corresponding shear stresses on the nearby cells in therapeutic applications like drug delivery.
	
	This study further supports the importance of the consideration of inter-bubble interactions in designing and optimizing applications using MBs. The majority of the applications using MBs in combination with ultrasound use polydisperse MB clusters. Our findings indicate that the sole consideration of isolated MB behavior as the basis to optimize their use in applications may be insufficient since the behavior of MBs can be altered dramatically due to constructive and destructive inter-bubble interactions. We have previously shown the crucial importance of considering the bubble-bubble interaction when characterizing the shell parameters through attenuation and sound speed measurements experiments \cite{70,71}

	In this work we have investigated the nonlinear behavior of a polydisperse MB cluster. We were able to classify three important regimes of interaction. The findings of this paper provides a fundamental insight on the behavior of interacting bubbles and can generate the building blocks of future analysis of larger bubble clusters and optimizing their behavior to enhance the relevant applications.
	
	\section{Acknowledgement}
	\justifying
	The work is supported by the Natural Sciences and Engineering Research Council of Canada (Discovery Grant RGPIN-2017-06496), NSERC and the Canadian Institutes of Health Research ( Collaborative Health Research Projects ) and the Terry Fox New Frontiers Program Project Grant in Ultrasound and MRI for Cancer Therapy (project \#1034). A. J. Sojahrood is supported by a CIHR Vanier Scholarship.

\end{document}